\documentclass[
    ,final            
  ]
  {aipproc}

\layoutstyle{6x9}

\begin{document}

\title{Charm Production in $\bar pA$-Collisions at the Charmonium 
Threshhold}


\author{L. Gerland}{
  address={SUBATECH, Laboratoire de Physique Subatomique et des
Technologies Associ\'ees, \\University of Nantes - IN2P3/CNRS - Ecole 
des Mines de Nantes, 4 rue Alfred Kastler, \\44072 Nantes, Cedex 03, 
France}
}



\begin{abstract} 
We discuss the production of charmonium states in antiproton-nucleus
collisions at the $\psi'$ threshold.  It is explained that measurements
in $\bar p A$ collisions will allow to get new information about the
strengths of the inelastic $J/\psi N$ and $\psi'N$ interaction, on the
production of $\Lambda_c$ and $\bar{D}$ in charmonium-nucleon
interactions and for the first time about the nondiagonal transitions
$\psi' N\to J/\psi N$. The inelastic $J/\psi$-nucleon cross section is
extracted from the comparison of hadron-nucleus collisions with
hadron-nucleon collisions.  Predictions for the ratio of $J/\psi$ to
$\psi'$ yields in antiproton-nucleus scatterings close to the threshold
of $\psi'$ production for different nuclear targets are presented. 
\end{abstract}

\maketitle


\section{Introduction}

This work is based on ref.~\cite{gerplb}. In this paper we make
predictions for antiproton-nucleus collisions at the $\psi'$ threshold.
This measurement will be possible at the future antiproton-nucleus
experiment at the GSI~\cite{gsi}. We demonstrate that in these
collisions the cross section for the nondiagonal transition $\psi'+N\to
J/\psi+N$ can be measured. We account for the dependence of the cross
sections on energy, and the dependence of the elastic cross section on
the momentum transfer.
                                                                                
The charmonium production in $\bar p A$ collisions at the $\psi'$
threshold is well suited to measure the genuine charmonium-nucleon cross
sections.  At higher energies formation time effects makes the
measurement of these cross sections more difficult~\cite{ger}. These
cross sections and the cross section for the analysis of charmonium
production data at SPS-energies~\cite{spieles,brat}. At collider
energies, i.e.\ at RHIC and LHC, the formation time effects will become
dominant and charmonium states will be produced only far outside of the
nuclei~\cite{ger2}. However, measurements of the genuine
charmonium-nucleon cross sections as well as the cross section for the
nondiagonal transition $\psi'+N\to J/\psi+N$ are also important at
collider energies for the evaluation of the interaction of charmonium
states with the produced secondary particles.

\section{Model Description and Results}

In the semiclassical Glauber-approximation the cross section to produce a
$\psi'$ in an antiproton-nucleus collision is
                                                                               
{\small \begin{eqnarray}
\sigma\left(\bar p + A\to \psi' \right)&=&2\pi\int db\cdot
b\,dz_1{n_p\over A}\rho(b,z_1)\sigma\left(\bar p +
p\to \psi' \right)exp\left(-\int\limits_{-\infty}^{z_1}
dz \sigma_{\bar p N inel}\rho(b,z)\right)\cr
& &\phantom{xxxxx}
\times exp\left(-\int\limits_{z_1}^{\infty}dz\sigma_{\psi'N
inel}\rho(b,z)\right)\quad.
\label{antipapsiprim}
\end{eqnarray}}
                                                                               
In this formula, $b$ is the impact parameter of the antiproton-nucleus
collision, $n_p$ is the number of protons in the nuclear target, $z_1$ is
the coordinate of the production point of the $\psi'$ in beam direction,
and $\rho$ is the nuclear density. $\sigma\left(\bar p + p\to
\psi'\right)$ is the cross section to produce a $\psi'$ in an
antiproton-proton collision. $\sigma_{\bar p N inel}$ is the inelastic
antiproton-nucleus collision.  $\sigma_{\psi' N inel}$ is the inelastic
$\psi'$-nucleon cross section.
                                                                               
All the factors in eq.~(\ref{antipapsiprim}) have a rather direct
interpretation. The first exponantial gives the probability to find an 
antiproton at the
coordinates $(b,z_1)$, which accounts for its absorption, and ${n_p\over
A}\rho(b,z_1)\sigma\left(\bar p + p\to \psi'\right)$ is the probability
to create a $\psi'$ at these coordinates. The factor ${n_p\over A}$ 
accounts
for the fact that close to the threshold the antiproton can produce a
$\psi'$ only in a collision with a proton but not with a neutron. The 
second gives the probability that the produced $\psi'$ has no inelastic 
collision in the nucleus, i.e.\ that it survives on the way out of the 
nucleus.
                                                                               
Similarly, in the semiclassical Glauber-approximation the cross section to
subsequently produce a $J/\psi$ in an antiproton-nucleus collision is
{\small \begin{eqnarray}
& &\sigma\left(\bar p + A\to J/\psi+X \right)=\cr
& &2\pi\int db\cdot b\,dz_1\,dz_2\,\theta(z_2-z_1){n_p\over
A}\rho(b,z_1)\sigma\left(\bar p + p\to
\psi' \right)exp\left(-\int\limits_{-\infty}^{z_1} dz \sigma_{\bar p N
inel}\rho(b,z)\right)\cr
& &\times exp\left(-\int\limits_{z_1}^{z_2}dz\sigma_{\psi' N
inel}\rho(b,z)\right)
 \sigma(\psi'+N\to\psi+N)\rho(b,z_2)
exp\left(-\int\limits_{z_2}^{\infty}dz\sigma_{\psi N
inel}\rho(b,z)\right)\quad .\cr
& &
\label{antipapsi}
\end{eqnarray}}

\begin{figure}
  \includegraphics[height=.35\textheight]{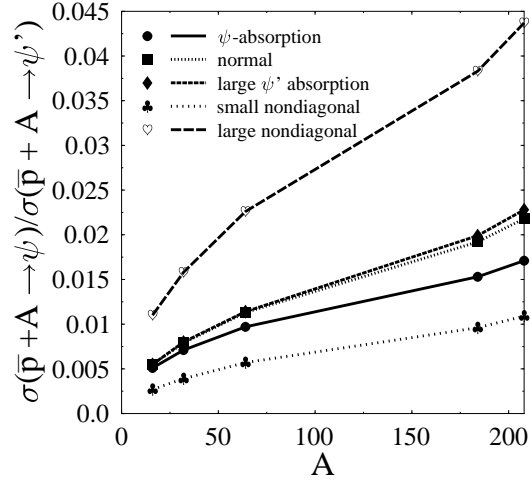}
  \caption{The ratio $\sigma(\bar p +A \to
\psi +nuclear fragments)/\sigma(\bar p +A \to
\psi' +nuclear fragments)$ is shown for 5 different sets of
parameters (see text for further details). Shown are the nuclear 
targets O, S, Cu, W, and Pb. The lines are just to guide the eye.}
  \label{antipa}
\end{figure}

In fig.~\ref{antipa} we used five sets of parameters. "normal" means
that the inelastic antiproton-nucleon cross section is $\sigma_{\bar p N
inel}=$50~mb, the inelastic cross section of the $\psi'$ is
$\sigma_{\psi'N inel}=$7.5~mb, the inelastic cross section of the
$J/\psi$ is $\sigma_{\psi N inel}=$0~mb, and the cross section for the
nondiagonal transition $\psi'+N\to J/\psi+N$ is
$\sigma(\psi'+N\to\psi+N)=$0.2~mb. The other sets differ by only
one of these parameters each:
\begin{itemize}
\item
In "$\psi$-absorption" $\sigma_{\psi N inel}=3.1$~mb.
\item
In "large $\psi'$ absorption" $\sigma_{\psi N inel}=15$~mb.
\item
In "small nondiagonal" $\sigma_{J/\psi N \to \psi'}=0.1$~mb.
\item
In "large nondiagonal" $\sigma_{J/\psi N \to \psi'}=0.4$~mb.
\end{itemize}
{\it One can see that the result depends much more strongly on the
nondiagonal cross section than on the absorption cross sections of the
$J/\psi$ and the $\psi'$. Therefore, this process is well suited to
measure the nondiagonal cross section.}

\begin{figure}
  \includegraphics[height=.33\textheight]{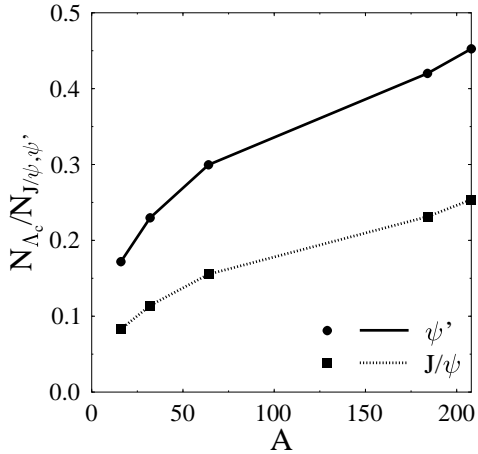}
  \caption{The ratio of the number of $\Lambda_c$ divided by the number 
of produced $J/\psi$ and $\psi'$ respectively states at the threshold of
$J/\psi$ and $\psi'$ production respectively. Shown are the nuclear
targets O, S, Cu, W, and Pb. The lines are just to guide the eye.}
  \label{lambdac}
\end{figure}

The cross section for the production of $\psi'$ that doesn't undergo an
inelastic rescattering is $\sigma\left(\bar p + A\to \psi' +\mbox{
nuclear fragments}\right)$ is given by eq.~(\ref{antipapsiprim}). The
cross section for the production of $\psi'$, whether they have
subsequent inelastic scatterings or not is given by
{\small \begin{eqnarray}
& &\sigma\left(\bar p + A\to \psi' \right)_{w/o inel}=\cr
& &2\pi\int db\cdot b\,dz_1{n_p\over A}\rho(b,z_1)\sigma\left(\bar p +
p\to \psi' \right)exp\left(-\int\limits_{-\infty}^{z_1}
dz \sigma_{\bar p inel}\rho(b,z)\right)\quad.
\label{antipapsiprimwo}
\end{eqnarray}}
Assuming that the $\Lambda_c$ channel is the only possible final state
in inelastic collisions (i.e.\ the $D\bar D$ channel as well as the
nondiagonal transition is neglected as a correction here), the fraction
of the initially produced $\psi'$ that ends up in the $\Lambda_c$
channel is
{\small \begin{equation}
{N_{\Lambda_c}\over N_{\psi'{initial}}}=1-{\sigma\left(\bar p +
A\to \psi' +nuclear fragments
\right)\over\sigma\left(\bar p + A\to \psi' +nuclear fragments
\right)_{w/o inel}}\quad.
\label{lamdacfrac}
\end{equation}}
Here we neglected the final state interactions of $\Lambda_c$ as they
may only effect the momentum distribution of $\Lambda_c$ since the
$\Lambda_c$ energy is below the threshold for the process
$p+\Lambda_c \to N + N + D$. For this reaction the $\Lambda_c$ would 
need an energy of 4.2~GeV in the rest frame of the proton, while it has 
in average less than 3~GeV. The change of the momentum distribution of
$\Lambda_c$ would provide unique information about the $\Lambda_c N$
interaction and could be a promising method for forming charmed
hypernuclei. Obviously eq.~\ref{lamdacfrac} is valid also for $\bar{D}$
production. The fraction for the $\psi'$ and the $J/\psi$ threshold is
depicted in fig.~\ref{lambdac}. 

{\it One can see that the result depends strongly on the
inelastic cross section of the $J/\psi$ and the $\psi'$. Therefore, this 
process is well suited to measure the inelastic cross section.}

\section{Conclusions}

It was shown that the future $\bar p A$-experiments at the GSI are well 
suited to measure genuine $J/\psi$ nucleon and $\psi'$ nucleon cross 
sections, i.e. the inelastic and the nondiagonal ($\psi' N\to J/\psi 
N$) cross sections.


\begin{theacknowledgments}
I am grateful to L.~Frankfurt and M.~Strikman for discussions  and I 
want to thank the organizers of the LEAP05 for giving me the chance to 
present my results.
\end{theacknowledgments}

\bibliographystyle{aipproc}   

\begin{thebibliography}{9}

\bibitem{gerplb}
L.~Gerland, L.~Frankfurt and M.~Strikman,
Phys.\ Lett.\ B {\bf 619}, 95 (2005)
[arXiv:nucl-th/0501074].
                                                                                
\bibitem{gsi} H.~Koch [PANDA Collaboration],
Nucl.\ Instrum.\ Meth.\ B {\bf 214} (2004) 50.
                                                                                
\bibitem{ger} L.~Gerland, L.~Frankfurt, M.~Strikman, H.~St\"ocker and
W.~Greiner,
Phys.\ Rev.\ Lett.\ {\bf 81}, 762 (1998)
[arXiv:nucl-th/9803034].\\
L.~Gerland, L.~Frankfurt, M.~Strikman, H.~St\"ocker and W.~Greiner,
Nucl.\ Phys.\ A {\bf 663}, 1019 (2000)
[arXiv:nucl-th/9908052].\\
L.~Gerland, L.~Frankfurt, M.~Strikman and H.~St\"ocker,
Phys.\ Rev.\ C {\bf 69}, 014904 (2004)
[arXiv:nucl-th/0307064].
                                                                                
\bibitem{spieles}
C.~Spieles, R.~Vogt, L.~Gerland, S.~A.~Bass, M.~Bleicher, H.~St\"ocker 
and
W.~Greiner,
Phys.\ Rev.\ C {\bf 60}, 054901 (1999)
[arXiv:hep-ph/9902337].
                                                                                
\bibitem{brat} E.~L.~Bratkovskaya, A.~P.~Kostyuk, W.~Cassing and
H.~St\"ocker,
Phys.\ Rev.\ C {\bf 69} (2004) 054903
[arXiv:nucl-th/0402042].
                                                                                
\bibitem{ger2} L.~Gerland, L.~Frankfurt, M.~Strikman, H.~St\"ocker and
W.~Greiner,
J.\ Phys.\ G {\bf 27} (2001) 695
[arXiv:nucl-th/0009008].
                                                                                
\end{thebibliography}

\end{document}